\documentclass[aps,twocolumn,showpacs]{revtex4}
\usepackage{graphicx}
\usepackage{dcolumn}% Align table columns on decimal point
\usepackage{amsmath}
\usepackage{amssymb}
\usepackage{bm}

\begin{document}

\title{Escape through a time-dependent hole in the doubling map}

\author{Andr\'e L.\ P.\ Livorati$^{1,2}$, Orestis Georgiou $^{2}$, Carl P. Dettmann$^{2}$ and Edson D.\ Leonel$^{3}$}

\affiliation{$^1$Instituto de F\'isica - IFUSP - Universidade de S\~ao
Paulo - USP  Rua do Mat\~ao, Tr.R 187 - Cidade Universit\'aria --
05314-970 -- S\~ao Paulo -- SP -- Brazil -- livorati@usp.br \\
$^2$ School of Mathematics - University of Bristol - Bristol BS8 1TW - United Kingdom \\ 
$^3$ Departamento de F\'isica -- UNESP -- Universidade Estadual Paulista -- Av. 24A, 1515 - Bela Vista -
13506-900 - Rio Claro - SP - Brazil \\
}

\begin{abstract}
We investigate the escape dynamics of the doubling map with a time-periodic
hole. Ulam's method was used to calculate the escape rate as a function of
the control parameters. We consider two cases, oscillating or breathing holes,
where the sides of the hole are moving in or out of phase respectively. We
find out that the escape rate is well described by the overlap of the hole
with its images, for holes centred at periodic orbits.
\end{abstract}

\maketitle

\section{Introduction}
\label{sec1}
A recent problem of interest among both physicists and mathematicians is the study of dynamical
systems with holes \cite{ref1}. Escape occurs when trajectories enter some predefined subset of the phase space called a hole.
This ``leaking'' of trajectories can happen in bounded \cite{ref2,ref5} 
as well in unbounded domain systems \cite{ref3,ref4,ref5a}. A natural observable allowing the study of the statistical
properties of this escape, in particular $\rho(n)$, the probability (given a suitable distribution
of initial conditions) that an orbit does not escape until a time $n$.
This raises the natural question of the decay rate of $\rho(n)$. The most important aspect of this analysis is that the escape rate is very 
sensitive to the system dynamics. For strongly chaotic systems the decay is typically exponential \cite{ref6}, while systems that
present mixed phase space (e.g. elliptic islands and a chaotic sea), the decay can be slower, presenting a mix of exponential 
with a power law \cite{ref7,ref7a}, or stretched exponential decay \cite{ref5a}. Indeed, when a non-exponential decay is observed
the dynamics would require a long range correlation, as for example a consequence of stickiness influence \cite{ref7}.
An equally important aspect is that the escape rate
can have a strong dependence on the position and size of the hole {\cite{ref5,ref7b}}. Applications of leaking systems can be found in
a great variety of fields, including plasmas \cite{app1,app2}, acoustics \cite{app3,app3a}, optics \cite{app4,app5}, fluids \cite{app6}, 
among others (see \cite{ref1} for a recent review).

While in most mathematical formulations of leaking systems the hole is static (and typically small relative to the phase space), 
in this paper we undertake a new approach and study escape through a time-dependent hole. 
Namely, we propose and investigate the escape properties of a chaotic leaking system where the hole position and hole 
size varies with time. Motivation for studying such problems can be traced back to the early 50$^{\prime}$s concerning Moshinsky$^{\prime}$s
shutter problem of ``diffraction in time" \cite{app7}. More recent applications can by found in quantum mechanics \cite{app7a,app7b}
and in atom-optics and ultra cold atoms experiments \cite{app8,app8a,app8b,app8c}. Further motivation for studying
time-dependent holes stems from chemical reactions and hydrodynamical flows (see for example the blinking vortex system \cite{app9}).

We restrict our investigations of time-dependent holes to the well studied and understood case of the open
doubling map (defined in the next section) \cite{ref5,ref8,ref9,ref10,ref11,ref12,ref13,ref14}.
What is particularly attractive about this map is that it is uniformly expanding with a uniform invariant density 
distribution and also has a well understood structure of periodic orbits due to the correspondence between the dynamics 
and the binary representation of phase space points. We aim to understand the role of these periodic orbits in the case of escape 
through a time-dependent hole, once they can play an important role in other dynamical systems \cite{ref15,ref16}. 
To this end, we present extensive numerical investigations and also construct accurate analytical predictions for $\rho(n)$.

The remainder of the paper is organized as follows: In Sec.\ref{sec2} we describe how the
time-dependent hole is introduced, and some properties concerning the escape
rate and the periodic orbits. The numerical and analytical results are shown
in Sec.\ref{sec3}. Finally some final remarks and conclusions are drawn in
Sec.\ref{sec4}.\

\section{The Mapping, properties and the time-dependent hole}
\label{sec2}

The dynamical system under study here is the doubling map modulo one, also known as the Bernoulli shift, represented by the
recurrence relation below
\begin{equation}
x_{n+1}=2x_n~~mod(1)~.
\label{eq1}
\end{equation}

\begin{figure}[htb]
\centering
\includegraphics[width=8cm]{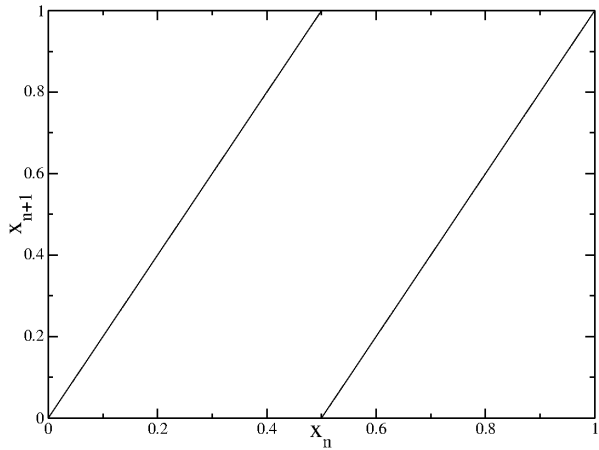}
\caption{{\it Phase space for the doubling map.}}
\label{fig1}  
\end{figure}

The phase space is shown in Fig.\ref{fig1}. Because of the simple nature of
the dynamics when we consider the 
binary notation \cite{ref5}, it is easy to categorize the dynamics based on the initial condition. If the initial 
condition is irrational, which are almost all points in the unit interval, the dynamics is non-periodic,
which follows directly from the definition of an irrational 
number as one with a non-repeating binary expansion. However, 
if $x_0$ is rational, the image of $x_0$ contains a finite number of distinct values within the interval $[0, 1)$ 
and the forward orbit of $x_0$ is eventually periodic, with period equal to the period of the binary expansion
of $x_0$. Particularly, if the initial condition is a rational number with a finite binary expansion of $k$ bits, 
then after $k$ iterations the iterates reach the fixed point $0$; if the initial condition is a rational number 
with a $k$-bit transient $(k\geq0)$ followed by a $p$-bit sequence $(p>1)$ that repeats itself infinitely, then after 
$k$ iterations the iterates reach a cycle of length $p$. Thus cycles of all lengths are possible \cite{ref5}. 
Another way of representing the periodic orbits is
\begin{equation}
x_0={q\over(2^p-1)}~,
\label{eq2}
\end{equation}
where $q\in \mathbb{Z}$ and $p$ is the period of the periodic orbit. So, for example, if we choose an initial condition with $q=8$ and $p=4$, 
it would be a period-4 orbit with dynamics evolving as $8/15\rightarrow1/15\rightarrow2/15\rightarrow4/15\rightarrow8/15$.

Once the main properties of the mapping are known, let us now introduce the escape in the dynamics, by considering a time-dependent
hole, i. e., a hole whose position and/or size are varying periodically. 

Defining the closed domain map as 
\begin{equation}
f:[0,1]\rightarrow[0,1]~,
\label{eq2a}
\end{equation} 
where $f$ is the application of the mapping in Eq.(\ref{eq1}). The open map is given by 
\begin{equation}
\hat{f}:[0,1] \backslash H_n \rightarrow [0,1]~,
\label{eq2b}
\end{equation} 
where $H_n\subset[0,1]$ is the hole at time $n$. Points within the hole are deemed to escape and are not considered further.

We set a mean fixed position for the hole to oscillate, $\bar{x}$, that could be in the neighbourhood of a short periodic orbit, 
or even the periodic orbit itself. Choosing the mean position around a periodic orbit allows us to compare the results we
obtain with the results already known in the literature for the fixed hole position \cite{ref7b,ref12,ref13,ref14}.

So, once the mean position is set up, we may define the hole size. Since, this value should vary with time, we can 
work with an average size of the hole, which we will name $\bar{h}$. So, two fixed positions for the hole to oscillate
were set. These positions represent the hole boundaries and we define them as the hole boundary at the right $h_r$ and the hole boundary 
at the left $h_l$, and they are set as
\begin{equation}
\left\{\begin{array}{ll}
h_r=\bar{x}+\bar{h}/2~\\
h_l=\bar{x}-\bar{h}/2~\\
\end{array}
\right.
\label{eq3}
\end{equation}
The expressions given in Eq.(\ref{eq3}), are saying that we have a hole with size $\bar{h}$, 
and its position is symmetric centred in $\bar{x}$.

\begin{figure}[htb]
\centering
\includegraphics[width=8cm]{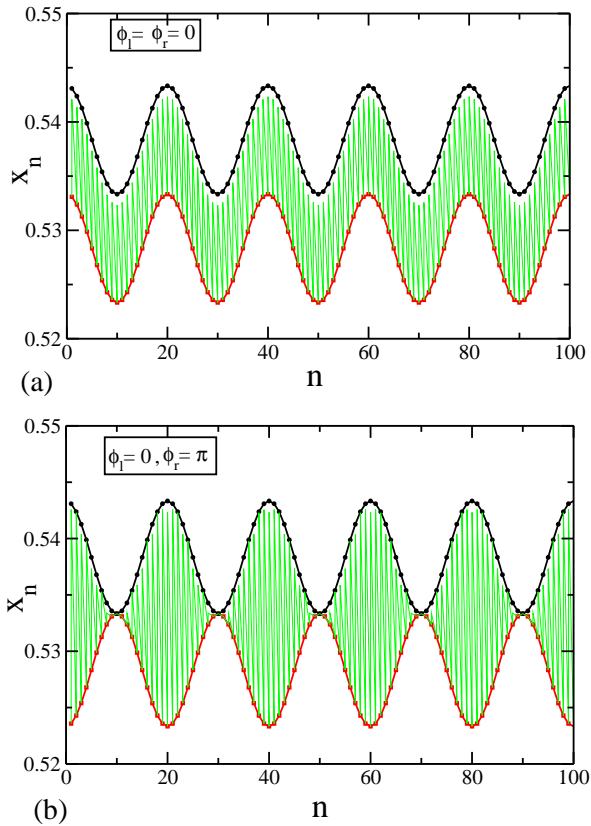}
\caption{Colour online: {\it Evolution of the time-dependent hole for two combinations of the initial phases for a value of $\tau=20$ and a mean
value around a period-4 orbit located in $(8/15)$. 
In (a) both, $h_l(n)$ and $h_r(n)$ are in phase with each other, so the size of the hole is kept constant during the dynamics 
and only its position is moving. And in (b), $h_l(n)$ and $h_r(n)$ are not in phase, so the size of the hole is varying, but
the average hole $\bar{h}$ is constant by period of oscillation. In this figure $h_l(n)$ is represented by the black 
line with bullets, and $h_r(n)$, setted as the red line with squares. The green dashed lines are the escape allowed region.}}
\label{fig2}  
\end{figure}

When we introduce the time dependence on the hole, we must deal with a discrete recurrence relation $(n)$, and not a 
continuous as time $(t)$. So, with a periodic oscillation, the expressions of Eq.(\ref{eq3})
can be written as

\begin{equation}
\left\{\begin{array}{ll}
h_r(n)=h_r+\epsilon\cos(\omega n+\phi_r)~\\
h_l(n)=h_l+\epsilon\cos(\omega n+\phi_l)~\\
\end{array}
\right.,
\label{eq4}
\end{equation}
where $\epsilon$ is the amplitude of oscillation of the holes, $\omega=(2\pi/\tau)$ is the frequency of oscillation and 
$\phi_l$ and $\phi_r$ are the initial phases of oscillation. The behaviour of each boundary of the hole, according $n$ evolves is 
illustrated in Fig.\ref{fig2}.

The value of the phases $\phi_l$ and $\phi_r$, in particular, whether they are equal or not, will influence the value of the 
amplitude of oscillation $\epsilon$, that may be chosen in order to keep the left and right boundaries of the hole
defined in Eq.(\ref{eq4}) in the $x$-axis domain. Figure \ref{fig2} shows how the hole would oscillate as $n$ evolves for
a mean position at a period-4 orbit. If $\phi_l$ and $\phi_r$ are in phase 
as shown in Fig.\ref{fig2}(a), the position of the hole is moving as $n$ evolves, but it remains with the same size; here we have 
no limit for the value of $\epsilon$, provided that is inside the domain of the doubling map. However, if $\phi_l$ and $\phi_r$ 
are not in phase, as shown in Fig.\ref{fig2}(b), the hole size is moving, in a breathing way. Here we have the limit 
breathing case that is $\epsilon\leq\bar{h}/2$. In this limit, we have a tangency between $h_l(n)$ and $h_r(n)$, when the period is complete, 
where the hole vanishes momentarily. If we go beyond this limit, there would be some prohibited regions for the escape,
and we are not considering this case in this paper.\

\section{Methods, Results and Discussions}
\label{sec3}
In this section we will present analytical and numerical methods to evaluate the escape through the time-dependent hole. We investigate how
the escape rate varies with the control parameters $\epsilon$, $\tau$ and the combination of initial phases $\phi_l$ and $\phi_r$.
We also make frequent comparison with the static hole case.

\subsection{Ulam's Method and Escape Rates}
We made use of Ulam's Method to calculate the escape rates for the time-dependent hole. Ulam's method is a
numerical scheme for approximating invariant densities of dynamical systems that can be made rigorous 
\cite{ref18,ref19,ref20,ref21,ref22}. 
The phase space is partitioned into connected sets and an inter-set transition matrix is computed from the dynamics; 
an approximate invariant density is read off as the leading left eigenvector of this matrix. When a hole in phase space 
is introduced, one instead searches for conditional invariant densities and their associated escape rates 
\cite{ref19,ref20,ref21,ref22,ref23,ref24,ref25,ref26}. In other words, we divide the space $X$ into a fine 
partition $X_i$, and assume that the probability of a transition from $i$ to $j$ is given by the proportion of $X_i$ that 
is mapped into $X_j$, that is

\begin{equation}
\rho_{ij}={|X_i\cap \hat{f}^{-1}(X_j)|\over|X_i|}~.
\label{eq5}
\end{equation}

If we consider the static hole case, given that the doubling map has exponential decay
of correlations, it seems clear that the survival probability should be exponential exponential, with a rate depending on the hole
position and size, as studied before in \cite{ref5,ref12,ref13,ref14}. For the same doubling map, the escape rate is

\begin{equation}
\gamma=-\lim_{n\rightarrow\infty}{1\over n}\ln\rho(n)~.
\label{eq5a}
\end{equation}

For a time-dependent hole, we may find an exponential decay related as suggested by Eq.(\ref{eq5a}), but the time dependence can also
have a superimposed periodic oscillation, as discussed in Sec.\ref{sec3}C.
 
\begin{figure}[htb]
\centering
\includegraphics[width=8cm]{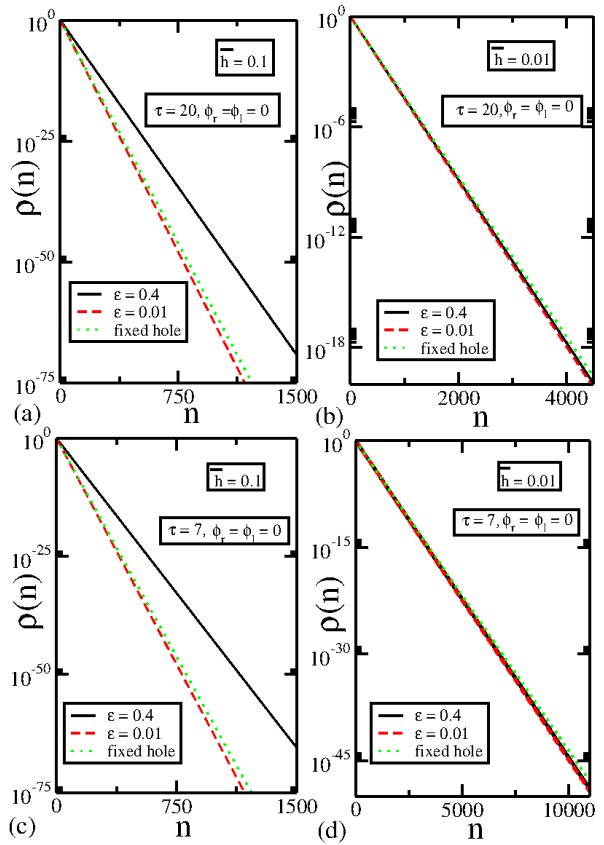}
\caption{Colour online: {\it Survival probability curves for different values of the average moving hole with equal initial phases. In (a) and
(b) we have $\tau=20$, and in (c) and (d) $\tau=7$. Also, we ranged the value of the amplitude of oscillation $\epsilon$ and
made a comparison with the fixed hole case. Depending of the combination of $\epsilon$ and $\tau$ we may have faster or slower 
escape.}}
\label{fig3}  
\end{figure}

We are considering the two different cases of initial phases of $\phi_l$
and $\phi_r$, as shown in Fig.\ref{fig2} in a separate way. Initial conditions used to calculate the survival 
probability for both kinds of holes were chosen equally split in the interval $[0,1]$.
Let us first address the case $\phi_l=\phi_r$ where only the hole position is moving and its size $\bar{h}$ is kept 
constant,
Fig. \ref{fig3} shows how these escape rates behave for some different average hole sizes, different values of amplitude 
of oscillation $\epsilon$ and different values of $\tau$.
For this figure, we decided to keep the mean value of the hole position $\bar{x}=8/15$. In later sections, we 
address other mean values of the hole $\bar{x}$. As expected in Figs.\ref{fig3}(a,c) with an average hole of $\bar{h}=0.1$, 
we have a much faster decay, than in Figs.\ref{fig3}(b,d), where the average hole is $\bar{h}=0.01$. However, one can also 
notice that depending on the combination of $\epsilon$ and $\tau$ parameters, we may have a faster or slower escape, as shown in 
Figs.\ref{fig3}(a,c), where for a bigger value of $\epsilon=0.4$ which basically contains the whole domain of the doubling map.
Also, one can notice that the labels of $\rho(n)$ axis in Fig. \ref{fig3} are very small. This precision is a result of the application
of Ulam method, which can be very accurate depending on the number of partitions.  
In Table \ref{Tab1} one may find the value of the escape rate for some
combinations of values of $\tau$ and $\epsilon$. 
 
\begin{table}[h]
%    \centering
%    \tiny
{
\begin{tabular}{|c|c|c|c|c|c|} \hline \hline
${\bar{h}}$&$\tau$&$\epsilon$&$\gamma$ \\
\hline
$0.1$&$7.0$&$0.4$&$0.1004784(1)$ \\
\hline
$0.1$&$7.0$&$0.01$&$0.1468975(5)$ \\
\hline
$0.1$&$20.0$&$0.4$&$0.1061757(6)$ \\
\hline
$0.1$&$20.0$&$0.01$&$0.1473164(5) $ \\
\hline
$0.1$&$0.0$&$fixed~hole$&$0.1418285(1)$ \\
\hline
$0.01$&$7.0$&$0.4$&$0.01024197(3)$ \\
\hline
$0.01$&$7.0$&$0.01$&$0.01030896(5)$ \\
\hline
$0.01$&$20.0$&$0.4$&$0.01020202(1)$ \\
\hline
$0.01$&$20.0$&$0.01$&$0.0103262(1)$ \\
\hline
$0.01$&$0.0$&$fixed~hole$&$ 0.0100119(1)$ \\ \hline
\hline
\end{tabular} }
\caption{Value of the escape rate $\gamma$ for a combination of $\epsilon$ and $\tau$ for two different hole sizes.}
\label{Tab1}
\end{table}

\begin{figure}[htb]
\centering
\includegraphics[width=8cm]{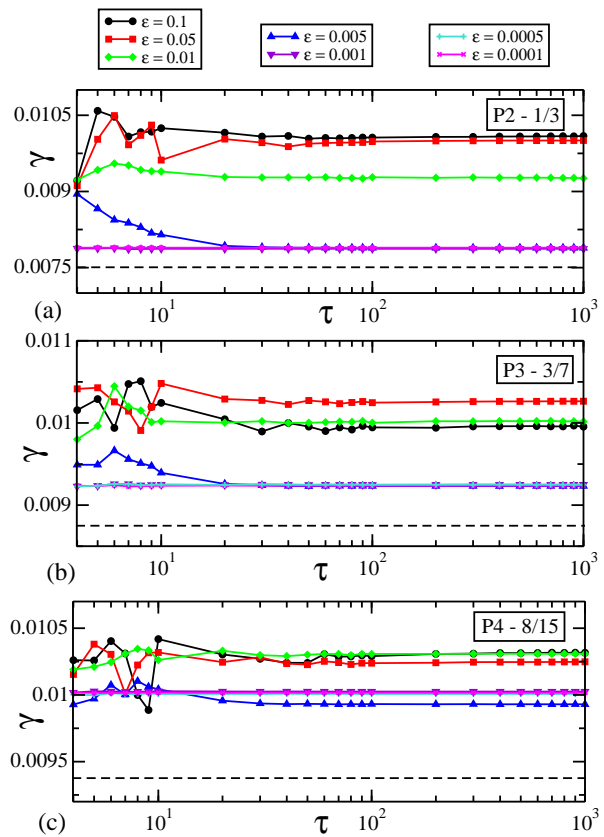}
\caption{Colour online: {\it Variation of the escape rate $\gamma$ as function of $\tau$ for several values of $\epsilon$. Here the average hole
is fixed at $\bar{h}=0.01$, and we considered three different periodic orbits for the mean hole position. In (a) $\bar{x}=1/3$, a 
period-2 orbit, in (b) $\bar{x}=3/7$, a period-3 orbit, and finally in (c) $\bar{x}=8/15$, a period-4 orbit. Notice 
that for high values of $\epsilon$, for a small $\tau$ regime, the escape rate varies a lot, and as $\tau$ increases it bend 
towards an almost constant regime. For the limit $\epsilon \rightarrow0$, the escape rate behaves closer as the one expected
for a fixed hole, for all values of $\tau$. The dashed lines represents the first order approximation of the fixed hole escape rate,
according Eq.(\ref{eq6}).}}
\label{fig4}  
\end{figure}

\begin{figure*}[htb]
\begin{center}
\centerline{\includegraphics[width=17cm,height=12.0cm]{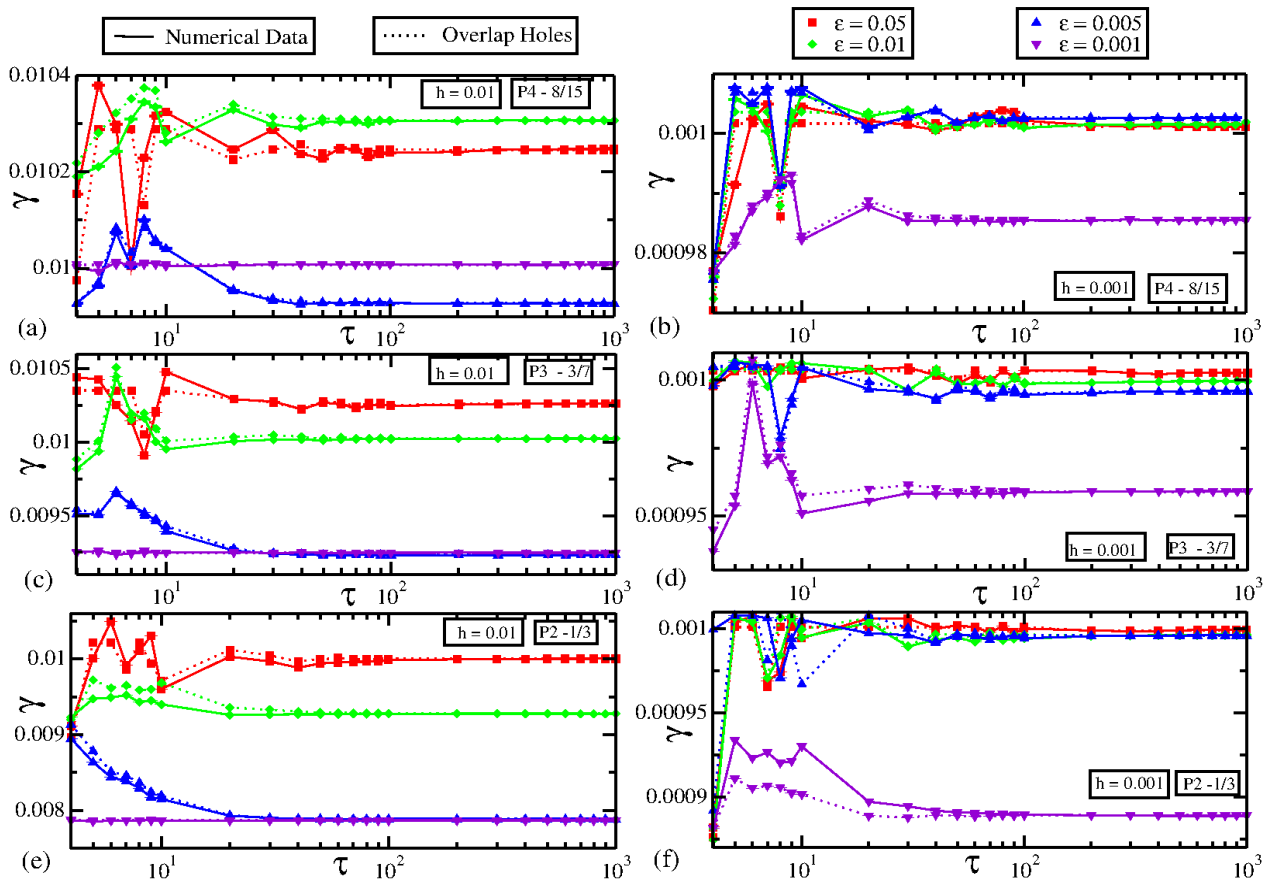}}
\end{center}
\caption{Colour online: {\it Comparison between the numerical data and the analytical approach for the overlap holes, given by Eqs.(\ref{eq7})
and (\ref{eq8}). In (a), (c) and (e) the average hole is $\bar{h}=0.01$, and in (b), (d) and (f) we have $\bar{h}=0.001$. 
Also, the numerical data is given by the full lines, and the analytical
approach is given by the dotted lines. Red squares represent $\epsilon=0.05$, green diamonds $\epsilon=0.01$, blue up triangles $\epsilon=0.005$ and
purple down triangles $\epsilon=0.001$. The matching is really good for high values of $\tau$, where the hole is moving slowly, but for
low values of $\tau$, where the hole is moving faster, there is still a gap between them.}}
\label{fig6}  
\end{figure*}

One can see, that the value of $\gamma$ is proportional to the average hole size, that would be roughly expected according to 
Refs.\cite{ref5,ref7b,ref11,ref14}, but there are also significant changes, depending of the combinations of the control 
parameters.
In order to understand better how the escape rate varies with $\epsilon$ and $\tau$, we plotted the value
of the escape rate $\gamma$ vs. $\tau$ for several values of $\epsilon$, as shown in Fig.\ref{fig4}. Here we keep the average
hole size $\bar{h}=0.01$ and consider at three different mean values for the hole to oscillate $(\bar{x})$ about three 
different periodic orbits. We see that for small values of $\tau$, there is a large 
variation in the escape rate for all values of the mean hole, considering some high values of $\epsilon$ (say, above 
$\epsilon=0.001$, which is $10$ percent of the average hole size). These fluctuations can be explained, once  
$\epsilon$ is big enough and $\tau$ small, the hole is moving up and down really fast as the dynamics evolves, and during this 
movement, it can intercept several different periodic orbits, which may be one of the explanation for the variation of the escape
rate. However, for high values of $\tau$, the escape rate stays almost constant. This should be expected, since for high values 
of $\tau$, the hole takes a time much longer then other time-scales in this problem to change its position. Also, if we 
look for the curves with small values of $\epsilon$, one can see that they 
stay in a constant regime for all values of $\tau$, as once $\epsilon \rightarrow0$, the moving hole starts to behave like
a fixed hole. Comparing the limit of $\epsilon \rightarrow0$ for Fig.\ref{fig4}(a,b,c), we can see that the plateau where
the escape rate establish itself changes, when we consider a different periodic orbit. Indeed, as shown in 
\cite{ref5,ref7b,ref13,ref14} the escape occurs faster through a hole
which contains long periodic orbits, and is slower if the hole contains short periodic orbits.
Recall that, according to the literature \cite{ref5,ref7b,ref14}, 
the escape rate through a small hole covering a short periodic orbit is approximately given by

\begin{equation}
\gamma=\bar{h}(1-\Lambda^{-1})~,
\label{eq6}
\end{equation}
where $\Lambda=2^p$ and $p$ is the period of the periodic orbit, represented by dashed lines for all the periodic orbits in
Fig.\ref{fig4}. We also observe higher order corrections to the expression
presented in Eq.(\ref{eq6}). These higher order corrections are specifically detailed in \cite{ref7b,ref14}, and we think that if they were 
taken into account together with Eq.(\ref{eq6}), there would be a good agreement between them.
Also, one could ask, by choosing a mean position for the hole to oscillate as an irrational number if there would be any different result.
We think that the results would be basically the same. The escape rate must be in somehow proportional to the hole
size and present itself as an exponential decay. Perhaps, a small difference would be the analytical treatment 
concerning the corrections related to periodic orbits, as the first order correction in Eq.(\ref{eq6}).

\subsection{Overlap Holes}
We now develop an analytic approach where both $\phi_l$ and $\phi_r$ are equal, as function of the control parameters 
$\tau$ and $\epsilon$, using the overlap of the periodic orbits with the
$p$ application of the mapping in Eq.(\ref{eq1}). The motivation for this kind of attempt came from Ref.\cite{ref8}, where an 
extensive analytical analysis is made concerning the escape rate on the doubling map. What the overlap hole approach do is 
basically a ``correction" of Eq.(\ref{eq6}), concerning the moving hole. Considering these overlaps, we can write the escape
rate as

\begin{equation}
\gamma_{oh}=\bar{h}\left(1-{1\over\tau}\sum_{i=0}^{\tau-1}{|f^p(H_i)\cap H_{i+p}|\over |f^p(H_i)|}\right)~,
\label{eq7}
\end{equation}
where the index $oh$ means the overlap holes, $H_i$ is the hole size in the $i$-th iteration, and $H_{i+p}$ is the hole 
size considered on the $i+p$-th iteration.

We make a comparison of the results obtained considering the numerical
simulations using Ulam's Method, and analytical approach by the formula
expressed in Eq.(\ref{eq7}), for three different periodic orbits of low period
and for two values of the average hole. The numerical data is represented by
the full lines, as the analytical approach is given by 
the dotted lines. Although, both data follows a similar behaviour, as we
increase the value of $\tau$, one can see in Fig.\ref{fig6} that there is still a gap between the numerical data and the 
analytical approach. We can attribute these gaps, to higher order corrections, once for a fixed hole the escape rate should follow
$\gamma=\bar{h}(1-\Lambda)+o(\bar{h})$, where the corrections may lie on the form $\bar{h}^2\ln{\bar{h}}$ \cite{ref14}.
The analytical data plotted in Fig.\ref{fig6} is hence adjusted according to
\begin{equation}
N[\gamma_{oh}(\tau)]=\gamma_{oh}(\tau)+\gamma_{n}(\infty)-\gamma_{oh}(\infty)~,
\label{eq8}
\end{equation}
where, $N[\gamma_{oh}(\tau)]$ is the normalized escape rate considering the overlap holes approach, $\gamma_{oh}(\tau)$, is the
analytical approach for the escape rate concerning the overlap holes
according Eq.(\ref{eq7}), $\gamma_{n}$, 
is the numerical escape rate obtained by Ulam's method. The argument $\tau\rightarrow\infty$, is taken along an average
between $\tau=100$ and $\tau=1000$, once the escape rate for this cases is almost constant.
So, with this correction, the higher order effects are taken into account, and the matching between the numerical and the 
analytical approach concerning the overlap holes in Eqs.(\ref{eq7}) and (\ref{eq8}) occurs. However, one can still see small
discrepancies for small values of $\tau$, where the hole is moving too fast. These still need further investigation.\

\subsection{Breathing Hole}
\begin{figure}[htb]
\centering
\includegraphics[width=9cm,height=10.0cm]{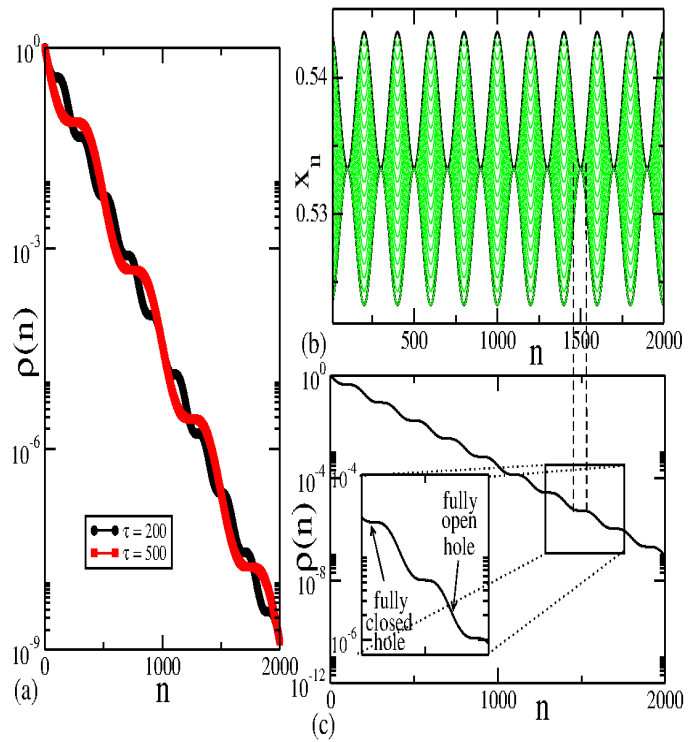}
\caption{Colour online: {\it Step-like decay behaviour for the probability curves when $\phi_r=0$ and $\phi_l=\pi$. We kept the average hole size constant
in $\bar{h}=0.01$ and $\epsilon=0.005$, which is $\epsilon=\bar{h}/2$. In (a) we have $\tau=200$ and $\tau=500$. One can 
notice that the step-like decays of the survival probability follows the period of oscillation in a exponential envelope. 
In (b) and (c) we have a comparison of the hole behaviour with the probability decays. When the hole is fully open, 
we have a faster escape, and when the hole is fully closed (tangency), we have a constant plateau, where no orbits are 
escaping. The zoom window in (c) shows better this step-like behaviour.}}
\label{fig7}  
\end{figure}

Now we address the case where $\phi_r=0$, and $\phi_l=\pi$, shown in Fig.\ref{fig2}(b), where the hole size is in 
constant change. Now, the hole is increasing and decreasing in a periodic way as $n$ evolves, in a breathing way, but the 
average hole size is the same over the period of oscillation. Figure \ref{fig7}(a) shows the escape rate curve for this kind of 
moving hole, for some values of $\tau$ for a hole centred in a mean position of $\bar{x}=8/14$. 
We can see that, in general it decays as an exponential envelope, but with a
peculiarity: it decays in steps. The step obeys the period of oscillation of
the moving hole, as shown in the comparison made in Figs.\ref{fig7}(b,c). The
steps can be basically explained by this comparison. Once the hole is
increasing and decreasing, the rate of orbits that escape through it varies
according to its size. So, when we have a tangency between both hole sides
$h_l$ and $h_r$, none of the orbits are escaping, then we have a constant
plateau of $\rho(n)$. On the other hand, when the hole is in its fully open size, we have a
faster escape. Here we used the value of $\epsilon=\bar{h}/2$, that is the
limit case for an instantaneous prohibited escape zone. If we had, a smaller value for
$\epsilon$, these step-like decays would be smoother, and in the limit that
$\epsilon\rightarrow0$, it would behave as a complete exponential curve decay.

An attempt for an analytical approach for the breathing hole can be made. We have that the hole is $h(n)=h_r(n)-h_l(n)$, 
where $h_r(n)$ and $h_l(n)$ are given by Eq.(\ref{eq4}), where the initial phases are $\phi_r=0$ and $\phi_l=\pi$. So, 
we may say that the hole size obeys
\begin{equation}
h(n)=\bar{h}+2\epsilon\cos(\omega n),
\label{eq9}
\end{equation}
where $\omega=2\pi/\tau$ and $\epsilon$ must not be bigger than $\bar{h}/2$.

Now, that the hole size is known we may propose the following expression where the decay of the survival 
probability as function of $n$, as shown in Fig.\ref{fig7}, is given by 
\begin{equation}
{d\rho\over dn}=-\rho\left\{\left[\bar{h}+2\epsilon\cos(\omega n)\right]\times P(noh)\right\}~,
\label{eq10}
\end{equation}
where $P(noh)$ is the probability of a point in the hole not overlapping with the hole on the $(n+p)$ iteration, 
just like we did for the the moving hole case, where again $p$ is the period of the periodic orbit where the hole is centred. 
This probability is given by
\begin{equation}
P(noh)={\bar{h}+2\epsilon\cos[\omega(n+p)]\over \bar{h}+2\epsilon\cos(\omega n)}\times{1\over\Lambda}~.
\label{eq11}
\end{equation}

Replacing the above expression and solving the separable equation, we have the following expression
\begin{equation}
\begin{array}{ll}
\rho(n)=\exp\{-\bar{h}n(1-\Lambda^{-1})~\\
-{2\bar{h}\epsilon\over\omega}[\sin(\omega n)+{\sin[\omega(n+p)]\over\Lambda}]\}~.
\end{array}
\label{eq12}
\end{equation}

In the above expression, the term $\bar{h}n(1-\Lambda^{-1})$ can be named as $\overline{\gamma(n)}$. Let us do an attempt to 
improve this expression making use of second order approximations. If we consider a fixed hole, the escape rate considering
second order effects \cite{ref8,ref14} can be given by
\begin{equation}
\gamma_{fixed}=h\left(1-\Lambda^{-1}\right)+{a_p}h^2\ln(h)~,
\label{eq13}
\end{equation}
where $a_p$ is a constant that may depend on the p-periodic orbits. In order, to find the $a_p$ value, we simulated for the several 
values of the fixed hole, and compared the numerical result of $\gamma=-\lim_{n\rightarrow\infty}{1\over n}\ln\rho(n)$, with 
the analytical approach of Eq.(\ref{eq13}), and found an average value for $a_p$, that is $a_{2}\approx1.812$, $a_3\approx2.055$
and $a_4\approx2.331$. We stress that the p-periodic orbits considered for the hole to be centred was the same ones used in the 
previous section.

In the breathing case, the hole size is in constant change as $n$ evolves, so we must assume that the escape rate is 
not constant either, as one can see in Fig.\ref{fig7}. So, according to Eqs.(\ref{eq9}) and (\ref{eq13}), we may set 
\begin{equation}
\gamma=h(n)\left(1-\Lambda^{-1}\right)+a_p{h(n)}^2\ln[h(n)].
\label{eq14}
\end{equation}

Once we have an average value for the hole as $\bar{h}$, we can consider also an average over the escape rate as 
\begin{equation}
\overline{\gamma}={\omega\over2\pi}\int_0^{2\pi\over\omega}\gamma dn~.
\label{eq15}
\end{equation}

Replacing Eq.(\ref{eq9}) and evaluating this average on Eq.(\ref{eq15}), one can obtain
\begin{equation}
\begin{array}{ll}
\overline{\gamma}=\bar{h}(1-\Lambda^{-1})+a_p\{\bar{h}[3\mu-\bar{h}(3\ln4)]~\\
-\epsilon^2(2\ln16)\}-2a_p(2\epsilon^2+{\bar{h}^2}){\ln(\bar{h}+\mu)\over2}~,
\end{array}
\label{eq16}
\end{equation}
where $\mu=\sqrt{-4\epsilon^2+{\bar{h}^2}}$. Now the step-like behaviour of $\rho(n)$ can be analytically expressed by the
combinations of Eqs.(\ref{eq12}) and (\ref{eq16}), where higher order effects are considered

\begin{equation}
\begin{array}{ll}
\rho(n)=\exp\{-\overline{\gamma}n-{2\bar{h}\epsilon\over\omega}\{(\sin(\omega n)
+{\sin[\omega(n+p)]\over\Lambda})\}~.
\end{array}
\label{eq17}
\end{equation}

\begin{figure*}[htb]
\begin{center}
\centerline{\includegraphics[width=17cm,height=12.0cm]{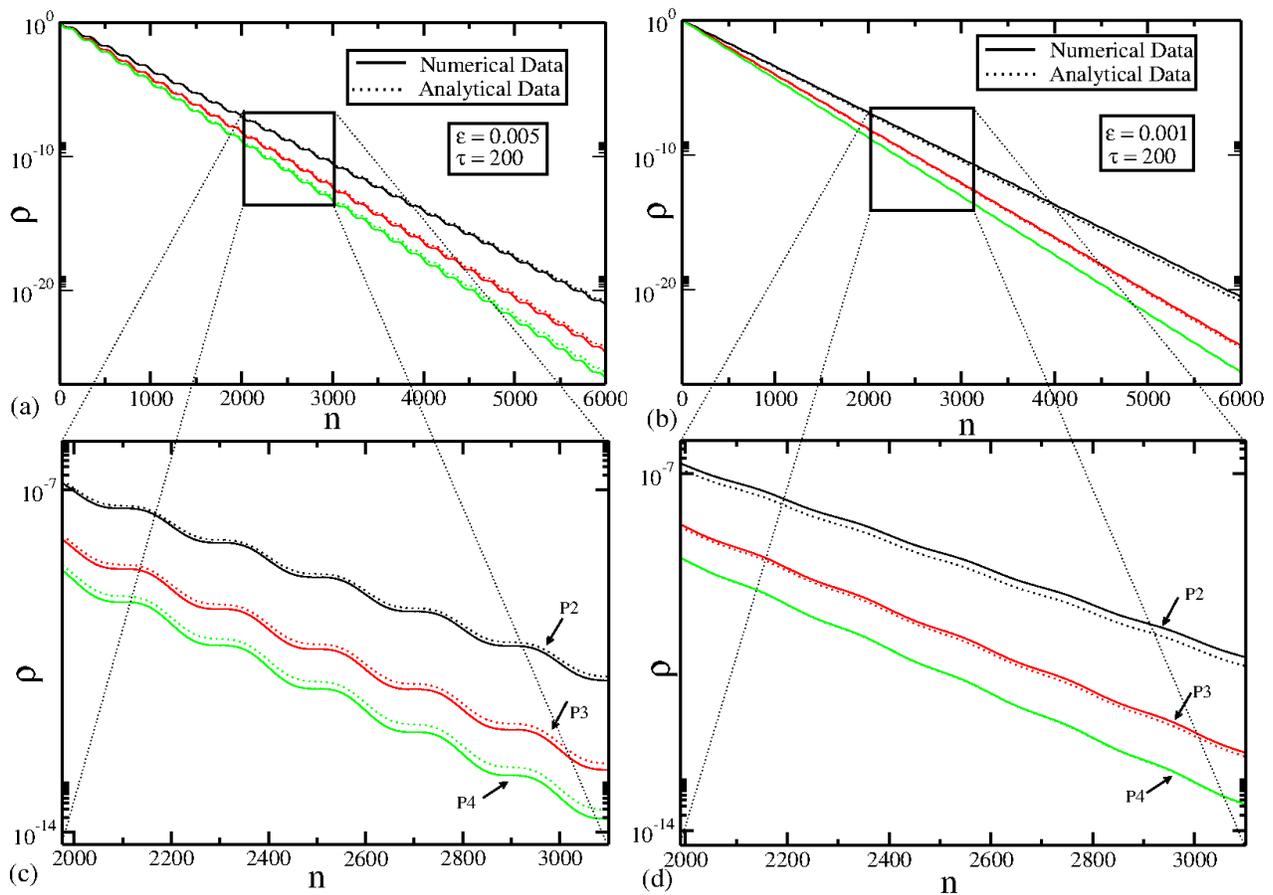}}
\end{center}
\caption{Colour online: {\it Comparison between the numerical step-like decays and the
analytical approach given by Eq.(\ref{eq17}) for $\tau=200$ and
$\bar{h}=0.01$. In (a) $\epsilon=0.005$ and in(b) $\epsilon=0.001$. One can
notice a remarkably good match between the numerical and the analytical data
in the amplifications in (c) and (d).}}
\label{fig8}  
\end{figure*}

Figure \ref{fig8} shows a comparison between the analytical (dotted lines) and the numerical data (full lines) 
for $\tau=200$, for three distinct periodic orbits, concerning a hole size $\bar{h}=0.01$. One can see that for the limit case 
$\epsilon=\bar{h}/2$ in Fig.\ref{fig8}(a), the step-like decay in the exponential envelope is present according to the 
hole period of oscillation and the analytical approach matches reasonably well. If we decrease the amplitude of oscillation,
to $\epsilon=0.001$ as shows Fig.\ref{fig8}(b), the survival probability curve presents a smooth behaviour concerning the 
step-like decays, and the exponential envelope is more dominant in the decay. For this case of smaller $\epsilon$ there 
is no more a instantaneous forbidden region in the hole evolution. However, concerning the zoom-in windows in 
Figs.\ref{fig8}(c,d), there is still a little gap between the numerical and analytical data. We believe, that this gap may
be due higher order effects, that may be introduced in Eq.(\ref{eq14}) in order to improve the analytical expression.

\section{Final Remarks and Conclusions}
\label{sec4}
We investigated the escape dynamics of the doubling map with a time-periodic hole with amplitude and period of 
oscillation, $\epsilon$ and $\tau$ respectively. We considered two
distinct ways for the hole to oscillate: {\it(i)} keeping the same size and changing its position, and {\it(ii)} breathing case.
This two kinds of hole are controlled by an initial phase $\phi_l$ and $\phi_r$ introduced in the time-dependent perturbation.

Using Ulam's method to calculate the probability of escape, we found that it is basically exponential, and for case 
{\it(i)} it depends on the value of $\tau$ and $\epsilon$. If we had a low $\tau$, the hole is moving really fast 
and we observe some fluctuations on the escape rate $\gamma$ vs. $\tau$ curves. If the hole is moving more slowly,
the escape rate is correspondingly more slowly varying with $\tau$. Also, for some high values of $\epsilon$, the hole can intercept many periodic
orbits, which can add even more fluctuations on the escape rate, and for $\epsilon\rightarrow0$, it reduces to a fixed hole. 
In an attempt to explain these fluctuations, we introduced an analytical approach related to overlap holes.
We observed that the numerical data and the analytical results, have excellent agreement if higher order effects are taken 
into account.
Considering case {\it(ii)}, we observed that the probability decays according 
a step-like function in an exponential envelope, which follows the value of
period of oscillation $\tau$ for the breathing hole.
We set up an analytical approach for the step-like decay also considering the probability of overlap holes, and it reasonably 
matches with the numerical data. Also, in the limit $\epsilon\rightarrow0$, the step-like is very smooth and the survival
probability can be expressed as an exponential law.

We emphasize that the control parameters strongly affect the escape rate, for the moving hole, considering
fast and slow moving hole, or the breathing case. As a next step, we would try to find the exactly higher order effects 
for the escape rate and improve the analytical expressions for both hole cases. Also, it would be interesting to see how the escape rate
would vary for periodically moving holes in more complicated systems, such as with mixed
phase space; and for non-periodic hole perturbations, for example random.

\acknowledgments
ALPL acknowledges CNPq and CAPES - Programa Ci\^encias sem Fronteiras - CsF
(0287-13-0) for financial support. EDL thanks
FAPESP (2012/23688-5), CNPq and CAPES, Brazilian agencies. ALPL also thanks the University of
Bristol for the kindly hospitality during his stay in UK. This research was
supported by resources supplied by the Center for Scientific Computing
(NCC/GridUNESP) of the S\~ao Paulo State University (UNESP). The authors are also grateful for 
fruitful discussions with Georgie Knight and Eduardo G. Altmann.

\end{document}